\documentstyle[twoside,fleqn]{article}
\newcommand{\bls}[1]{\renewcommand{\baselinestretch}{#1}}
\def\onecol{\onecolumn \mathindent=2em}
\def\twocol{\twocolumn \mathindent=1em}
\def\noi{\noindent}
\def\n{\noindent}
\def\nq{\hspace{-1em}}
\def\nqq{\hspace{-2em}}
\def\nhq{\hspace{-0.5em}}
\def\nhh{\hspace{-0.3em}}
\def\cm{\hspace{1cm}}
\def\inch{\hspace{1in}}
\def\yy{\\[5pt]}
\def\al{&\nhq}
\def\eql{\al =\al}
\def\nnn{\nonumber\\ \lal }
\def\lal{&&\nqq {}}               % left alignment
\def\nn{\nonumber\\ {}}

\newcommand{\gs}{\sigma}
\newcommand{\gt}{\theta}
\def\vp{\varphi}
\def\b{\beta}
\def\k{\kappa}
\def\L{\Lambda}
\def\sr{\sqrt}
\def\en{\eqno}
\def\DAL{\raisebox{-1.6pt}{\large $\Box$}\,}
\def\G{\Gamma}
\def\ve{\varepsilon}
\def\ep{\epsilon}
\def\l{\lambda}
\def\d{\delta}
\def\z{\zeta}
\def\pt{\partial}
\def\lln{{\rm ln}}
\def\const{{\rm const}}
\def\eq{Eq.\,}
\def\eqs{Eqs.\,}
\newcommand{\arctg}{\arctan}
\newcommand{\tg}{\tan}
\newcommand{\ch}{\cosh}
\newcommand{\sh}{\sinh}
\newcommand{\th}{\tanh}
\newcommand{\cth}{\coth}
\newcommand{\e}{{\rm e}}
\newcommand{\Arch}{\mathop{\rm Arcosh}\nolimits}

\newcommand{\heads}[2]{\markboth{\protect\small\it #1}{\protect\small\it #2}}
\newcommand{\Acknow}[1]{\subsection*{Acknowledgement} #1}
\newcommand{\Title}[1]{\noindent {\Large #1} \\}
\newcommand{\Author}[2]{\noindent{\large\bf #1}\\[2ex]\noindent{\it #2}\\}
\newcommand{\Abstract}[1]{\vskip 2mm \begin{center}
        \parbox{16.4cm}{\small\noi #1} \end{center}\medskip}
\newcommand{\PACS}[1]{\begin{center}{\small PACS: #1}\end{center}}
\newcommand{\email}[2]{\footnotetext[#1]{e-mail: #2}}

\def\beq{\begin{equation}}
\def\eeq{\end{equation}}
\def\bear{\begin{eqnarray}}
\def\ear{\end{eqnarray}}
\def\bearr{\begin{eqnarray} \lal}
\def\earn{\nonumber \end{eqnarray}}

\def\LAP{\raisebox{-1.6pt}{\large $\triangle$}\,}
\def\vp{\varphi}
\def\half{\frac{1}{2}}
\def\Arch{{\rm Arch}}
\def\Arth{{\rm Arth}}
\def\Arcth{{\rm Arcth}}
\def\arctg{{\rm arctg}}
\def\newpic#1{%
   \def\emline##1##2##3##4##5##6{%
      \put(##1,##2){\special{em:point #1##3}}%
      \put(##4,##5){\special{em:point #1##6}}%
      \special{em:line #1##3,#1##6}}}
\newpic{}
\bls{1.03}
\heads{D.Yu.Shabalkin}
{
On the Method of Effective Nonlinear Sigma Model.
}

\begin{document}
\twocolumn[

\bigskip

\Title{ON THE METHOD OF EFFECTIVE NONLINEAR SIGMA MODEL \yy
IN PLANE- AND AXIALLY-SYMMETRIC VACUUM SPACETIMES}

\Author{D.Yu.Shabalkin}
{Ulyanovsk State University,
42 Leo Tolstoy St., Ulyanovsk 432700, Russia}

\Abstract
{
In present article effective nonlinear sigma model (NSM) is considered.
Einstein equation solution, corresponded to the chiral fields
determined by functional parameter method, are presented.
Effective NSM of stationary axially-symmetric gravitational field is
constructed. Motion equations are solved exactly by functional parameter
method. Einstein equations solution are constructed.
For particular dependences of functional parameter graphics of solutions
are presented. Metric coefficients behaviour is shown to be similar as
functional parameter one.
}

] %%%%%%%%%%%%%%%%%%%%%%%%%%%%%%%%%%%%%%%%%%%%%%%%%

\email 1 {sd@themp.univ.simbirsk.su, sd@sv.uven.ru}

\section{Introduction}

In the present article the following block-diagonal metrics
are considered

\beq\label{basic}
ds^2=g_{ab}dx^adx^b+g_{\mu\nu}dx^{\mu}dx^{\nu}.
\eeq
Metric coefficients are supposed to be depended on $x^\mu,\,x^\nu$.
Block $g_{\mu\nu}$ is sure to be trans\-for\-med to conformal-plane form.
Then metrics shall take the form
$ds^2=g_{ab}dx^adx^b-f \eta_{\mu\nu}dx^{\mu}dx^{\nu},\,
\eta_{\mu\nu}={\rm diag}(1,\epsilon ),\,\epsilon ={\rm sign}\det g_{ab},\,f$
depends on $x^{\mu},x^{\nu}$ only. Signature $(-,-,-,+)$ is accepted.
Choosing $\{x^{\mu},x^{\nu}\}=\{z,t\}, \{x^a,x^b\}=\{x,y\}$ leads to plane-symmetric
case, $\{x^{\mu},x^{\nu}\}=\{r,z\}, \{x^a,x^b\}=\{t,\phi\}$ corresponds to
stationary axially-symmetric spacetimes.

Spacetimes, corresponding to (\ref{basic}) was in\-ten\-si\-vely discussed from
different point of view during last 10-20 years.

Inverse scattering method (ISM) technique was applied for Einstein equations
solution at first in the work \cite{BelZakh}. Cosmological applications was
based on solutions mentioned above \cite{BelVAWE}, \cite{OliVer},
\cite{GarVerd}. Nonlinear transformations put in the base of ISM make
impossible, in general, the construction of solution, possessing some definite
demanded properties. Analysis of solution obtained is rather difficult due to
the same reasons. Some ways to this problem solution were suggested in
work \cite{Alex81}, \cite{Letel84}, where ISM was presented in modified form.

Other avenue of spacetimes (\ref{basic}) investigation is connected with
some special solutions construction, such as Gowdy cosmological
solutions \cite{Gowd71}, \cite{Gowd74}. The solution obtaining process
was aimed at inhomogeneous universe construction.

Method suggested in \cite{MatMis67} devoted to symmetry investigation of
(\ref{basic}) was applied for exact solution obtaining in works
\cite{CheMus}, \cite{Ch97m}, \cite{ShCh97}, \cite{ensm-gc97}.
Some ad\-van\-tage of this method are represented here.

This way is based on the possibility of construction of ef\-fec\-tive
nonlinear sigma model (NSM)
\beq\label{Lnsm}
{\cal L}_{NSM}= \sqrt{|g|} \lbrace\half
 h_{AB}(\varphi) \varphi^A_{,i} \varphi^B_{,k} g^{ik} \rbrace,
\eeq
motion equations of which
\beq\label{DEnsm}
\frac{1}{\sqrt{|g|}}\partial_i(\sqrt {|g|} \varphi_{A}^{,i})-
 \Gamma_{C,AB} \varphi_{,i}^B \varphi_{,k}^C g^{ik}=0,
\eeq
appear to be equivalent to vacuum Einstein equations
\beq\label{VEE}
R_{ik}=0.
\eeq

In work \cite{ensm-gc97} dynamic equations of effective NSM
(\ref{Lnsm}), (\ref{DEnsm}) have been solved by the method of
functional parameter. The problem of vacuum Einstein equations
solution construction hasn't been taken into consideration.
Here we will use the method mentioned above for construction and
solutions analysis of (\ref{VEE}).

Consider two important classes of spacetimes, derived from (\ref{basic})
by appropriated coordinate choosing: plane-symmetric and axially-symmetric.

\section{Exact solutions in case of plane-symmetric spacetime}

Consider a spacetime
\beq\label{ps}
ds^2=A dx^2+2B dx dy+C dy^2-D[dz^2-dt^2],
\eeq
$\sqrt g=\alpha D,\alpha^2\equiv{AC-B^2};$ functions $A,B,C,D$
depend only on $z,t$.

Vacuum Einstein equation $R_{ik}=0$, having been shown \cite{CheMus},
are equivalent to effective NSM equations
\bear\label{NSMequatPS}
\lal\DAL e^\psi=0,\nnn
 \DAL\theta+(\psi_{,z}\theta_{,z}-\psi_{,t}\theta_{,t})-{1\over2}(\chi^{\enspace 2}_{,z}-
\chi^{\enspace 2}_{,t}) \sh2\theta=0,\nnn
 \DAL\chi+(\psi_{,z}\chi_{,z}-\psi_{,t}\chi_{,t})\nnn
\cm+2(\theta_{,z}\chi_{,z}-\theta_{,t}\chi_{,t})  \cth\theta=0,\\
\lal \DAL\phi-{1\over2}(\psi^{\enspace 2}_{,z}-\psi^{\enspace 2}_{,t})+
{1\over2}(\theta^{\enspace 2}_{,z}-\theta^{\enspace 2}_{,t})+\nnn
\cm+{1\over2}(\chi^{\enspace 2}_{,z}-\chi^{\enspace 2}_{,t})\sh^2\theta=0,\nonumber
\ear
$\DAL\equiv\partial_{zz}-\partial_{tt},$ \cite{CheMus}, if
metric coefficients are connected with chiral fields as follows
\bear\label{MetCoef}
A\eql-\e^\psi(\cos\chi \sh\theta+\ch\theta),\nn
B\eql\e^\psi \sin\chi\sh\theta,\nn
C\eql\e^\psi(\cos\chi\sh\theta-\ch\theta),\\
D\eql\e^\phi.\nonumber
\ear

Equations (\ref{NSMequatPS}) have been solved by functional parameter
method, suggested in \cite{ensm-gc97}. In this method
\bear\label{PSclassI}
\psi=\ln\xi,&&\theta=\theta(\xi),\nn
\chi=\chi(\xi)&&\phi=\phi(\xi),\\
&&\xi=\xi(z,t).\nonumber
\ear
According to the first equation of (\ref{NSMequatPS})\linebreak[1]
\mbox{$\DAL\xi(z,t)=0$}.
This way lead to the following solutions of the (\ref{NSMequatPS}).
\bear\label{FS}
\lal\psi=\ln\xi,\, \DAL\xi(z,t)=0\nnn
\theta=\Arch\left\{k/2\left(\left(\xi/\xi_0\right)^a+
\left(\xi/\xi_0\right)^{-a}\right)\right\}\nnn
\chi_{\pm}=\arctg\chi_0\\
\lal\pm\arctg\left\{|a/c|
\displaystyle\frac{\left(\xi/\xi_0\right)^{2a}+1}
{\left(\xi/\xi_0\right)^{2a}-1}\right\}\nnn
\phi=\phi_0+\phi_1\xi-\frac{a^2-1}{2}\xi,\, k=\sqrt{1+c^2/a^2}.\nonumber
\ear
Here $a,c$ are arbitrary constants.

Other solutions family, suggested in \cite{ensm-gc97}, connected with
supposition
\bear\label{PSclassII}
\psi=\const,&&\theta=\theta(\xi),\nn
\chi=\chi(\xi),&&\phi=\phi(\xi),\\
&&\xi=\xi(z,t).\nonumber
\ear

In this case equations take the form
\bear\label{FS2}
\lal\theta=\pm\Arch\left[ \frac{k}{2} \ch(a(\xi-\xi_0)\right]\nnn
\chi=\arctg\chi_0\pm \arctg\left[\displaystyle\left|\frac{a}{c}\right|
\th\left\{a(\xi-\xi_0)\right\}\right]\\
\lal\phi=\phi_0+\phi_1\xi+\frac{1}{4}a^2\xi^2.\nonumber
\ear

Metrics coefficients, reconstructed by (\ref{MetCoef})
for the case (\ref{FS}) may be represented in such a way
\bear\label{ABC}
A\eql-\xi\left(\left|\displaystyle\frac{1\mp \chi_0 G(\xi)}{\alpha}\right|\sqrt{\displaystyle\frac{F^2(\xi)-1}{G^2(\xi)+1}}+F(\xi) \right),\nn
B\eql\xi\left(\left|\displaystyle\frac{\chi_0\pm G(\xi)}{\alpha}\right| \sqrt{\displaystyle\frac{F^2(\xi)-1}{G^2(\xi)+1}}\right),\\
C\eql\xi\left(\left|\displaystyle\frac{1\mp \chi_0 G(\xi)}{\alpha}\right|\sqrt{\displaystyle\frac{F^2(\xi)-1}{G^2(\xi)+1}}-F(\xi) \right),\nonumber
\ear
\bear
F(\xi)\eql\frac{k}{2}\left(\left(\xi/\xi_0\right)^a+
\left(\xi/\xi_0\right)^{-a}\right),\nn
G(\xi)\eql\displaystyle\left|\frac{a}{c}\right|\frac{\left(\xi/\xi_0\right]^{2a}+1}
{\left(\xi/\xi_0\right)^{2a}-1},\,\alpha^2=1+\chi_0^2.\nonumber
\ear

The form of metric coefficient, associated with the second solutions family,
coincidence with (\ref{ABC}), but dependencies $F(\xi)$ and $G(\xi)$
are following.
\bear
F(\xi)\eql\frac{k}{2}\ch(a(\xi-\xi_0),\\
G(\xi)\eql\displaystyle\frac{|a|}{|c|}\th\left\{a(\xi-\xi_0)\right\}
\ear
Formally, the only restriction to the form of parameter-function $\xi(z,t)$
is $\xi>0$. In this case both chiral fields and metric coefficients will be
real functions for all $z,t\in{\bf R}$.

\begin{figure}[t]
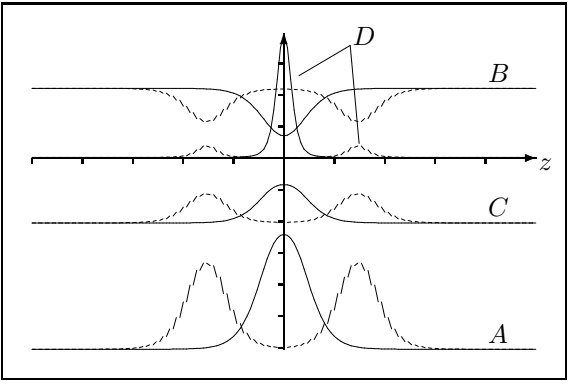


\unitlength=1mm


\caption{Gravitation fields at $t=0$ and $t=3$
         $a=2,c=1,\xi_0=10,\chi_0=10$.}\label{McPs}
\end{figure}

For instance of solution construction consider the function
\beq\label{soliton}
\begin{array}{l}
\xi(z,t)=\exp\left(\ch^{-2}(z-t)\right)\\
+\exp\left(\ch^{-2}(z+t)\right)
\end{array}
\eeq
for the case (\ref{PSclassI})-(\ref{FS}). It's known that dependence
(\ref{soliton}) describes solitons, moving to the opposite direction
at the light speed.
Graphics of the solution are represented on figure \ref{McPs}.
The forms of the metric coefficients at $t=0$ are shown by solid line
and by dotted line at $t=3$.

Solution for $g_{11}=A(z,t)$, $g_{12}=B(z,t)$, $g_{22}=C(z,t)$,
\mbox{$g_{33}=D(z,t)$}
conservative the soliton-like character. Impulse corresponding to
$D(z,t)$ influence the considerable amplitude decries since $t=0$ to $t=1$,
surpassing expected one according to (\ref{soliton}).
During the further evolution form of metric coefficients experience no
changes.

Possibility of $\xi(z,t)$ choice allows to construct solutions of the Einstein
vacuum equation possessing the determined properties. This is a advantage
of functional parameter method over Belinskii-Zakharov method \cite{BelZakh},
which set no direct connection between seed solution and solution obtained.

\section{Effective nonlinear sigma model of axially-symmetric spacetime}

In order to apply this method to axially-symmetric spacetimes analysis
it is necessary to construct effective NSM in this case.

Consider spacetime in the following form
\bear\label{as}
ds^2\eql\e^{2\nu(z,r)}dt^2-\e^{2\mu(z,r)}(dz^2+dr^2)\\
\lal-\e^{2\rho(z,r)}(d\phi-\omega(z,r) dt)^2.\nonumber
\ear

Connection between effective chiral fields and metric coefficients should
be defined by the way allowing

i)representation of the Legrangian of gravitation field (\ref{as})
$
{\cal L}_G=\sqrt{-g}g^{ik}\left(\Gamma^{l}_{km}\Gamma^{m}_{il}+\Gamma^{l}_{ik}\Gamma^{m}_{lm}\right)
$
as effective NSM Legrangian (\ref{Lnsm});

ii)transformation of Einstein equations to motion equations of the
corresponding effective NSM.

These conditions are satisfied both in the following case.

Relation between chiral fields and metrics coefficients
is determined in such a way
\beq\label{ASMetCoef}
\begin{array}{l}
\e^{2\nu}=2\e^\psi\ch\theta,
\omega=-\displaystyle\frac{\sh\chi}{\ch\chi-\cth\theta}\\
\e^{2\mu}=\e^\phi,
\e^{2\rho}=-\e^\psi (\ch\chi \sh\theta-\ch\theta).
\end{array}
\eeq
Fields are defined on space $dS^2=dz^2+dr^2$.
Intrinsic NSM space is chosen in form
\beq\label{ASchispace}
h_{IK}={e^\psi}\left(\matrix {-1&0&0&1\cr
                                 0&-1&0&0\cr
                                 0&0&\sh^2\theta&0\cr
                                 1&0&0&0\cr}\right),
\eeq
Dynamical equations in this case will be
\bear\label{NSMequatAS}
\lal\LAP\e^\psi=0,\nnn
 \LAP\theta+(\psi_3\theta_{,z}-\psi_{,r}\theta_{,r})+{1\over2}
(\chi^{\enspace 2}_{,z}-\chi^{\enspace 2}_{,r})   \sh 2\theta=0,\nnn
\LAP\chi+(\psi_{,z}\chi_{,z}-\psi_{,r}\chi_{,r})\nnn
\cm+2(\theta_{,z}\chi_{,z}-\theta_{,r}\chi_{,r})\cth\theta=0,\\
\lal \LAP\phi-{1\over2}(\psi^{\enspace 2}_{,z}-\psi^{\enspace 2}_{,r})+
{1\over2}(\theta^{\enspace 2}_{,z}-\theta^{\enspace 2}_{,r})\nnn
\cm-{1\over2}(\chi^{\enspace 2}_{,z}-\chi^{\enspace 2}_{,r})\sh^2\theta=0,\nonumber
\ear
$\LAP\equiv\partial_{zz}+\partial_{rr}.$

\section{Exact solutions}

Solution will be south by functional parameter method. According to its
ideology consider
\bear\label{ASclassI}
\psi=\ln \xi(z,r)&&\theta=\theta(\xi),\nn
\chi=\chi(\xi)&&\phi=\phi(\xi).
\ear

The ordinary differential equation set will be achieved

\bear\label{ODUas}
\lal\ddot\theta+\frac{1}{\xi}\dot\theta+\frac{1}{2}\dot\chi^2\sh2\theta= 0,\nnn
\ddot\chi+\frac{1}{\xi}\dot\chi+{2}\dot\chi\dot\theta\cth\theta= 0,\\
\lal\ddot\phi-\frac{1}{2\xi^2}+\frac{1}{2}\dot\theta^2+\frac{1}{4}\dot\chi^2\sh^2\theta = 0\nonumber
\ear
Here and henceforth
$\dot \vp_i\equiv \displaystyle{\partial \vp_i\over\partial \xi}$
Solutions of (\ref{ODUas}) depending on $\xi$ will posses form
\bear\label{FSas}
\theta\eql\Arch\left\{k \sin\left|\ln\left(\xi_0/\xi\right)^{a}\right|\right\}\nn
\chi\eql\Arth\chi_0\pm\frac{a}{\sqrt{c^2-a^2}}\nnn
\times\arctg\left\{\pm\frac{c}{\sqrt{c^2-a^2}}
\tg\left|\ln\left(\xi_0/\xi\right)^{a}\right|\right\}\\
\phi\eql\phi_0+\phi_1\xi+\frac{a^2-1}{2}(\ln\xi+1)-
   á \xi\displaystyle\int\limits^\xi\frac{\chi(\xi')}{\xi'^2}d\xi',\nn
k\eql\sqrt{1+c^2/a^2},\,a^2<c^2,\,c>0,\, a<0.\nonumber
\ear

Metric coefficients may be obtained by (\ref{ASMetCoef}).
\bear\label{MCas}
\lal\e^{2\nu}=2\xi F(\xi)\nnn
\e^{2\rho}=\xi F(\xi)-\xi\left|\displaystyle\frac{1\pm\chi_0\th G(\xi)}{\alpha}\right|\nn
\lal\times\sqrt{\displaystyle\frac{F^2(\xi)-1}{1-\th^2G(\xi)}},\nnn
\omega=-(\chi_0\pm\th G(\xi))\\
\lal\times\left(|1\pm\chi_0 \th G(\xi)|-|\alpha|F(\xi)
\sqrt{\frac{1-\th^2 G(\xi)}{F^2(\xi)-1}}\right)^{-1}.\nonumber
\ear

\bear
F(\xi)\eql k \sin\left|\ln\left(\xi_0/\xi\right)^{a}\right|,\nn
G(\xi)\eql\frac{a}{c^2-a^2}\arctg\left[\pm\frac{c}{\sqrt{c^2-a^2}}
\tg\left|\ln\left(\xi_0/\xi\right)^{a}\right|\right],\nn
\alpha^2\eql (1-\chi_0)^2.\nonumber
\ear

There are some conditions for function-pa\-ra\-me\-ter, necessary to existence
chiral fields and metric coefficients.

\bear\label{condxi}
\lal\frac{\arcsin k^{-1}+2\pi n}{|a|}\le |\ln \frac{\xi(z,r)}{\xi_0}|<
\frac{\pi(2 n+\half)}{|a|},\nnn
\ch\chi(\xi)<\cth\theta(\xi).
\ear

The first inequality of (\ref{condxi}) can't be solved analytically.
So it's impossible
to determine the conditions of restrictions for $\xi(z,r)$ in general case
with arbitrary constants. As soon as constants-parameters are defined
inequalities may be solved. Solutions of (\ref{condxi}) determine
$\xi(z,r)$ value set.

\begin{figure}[b]
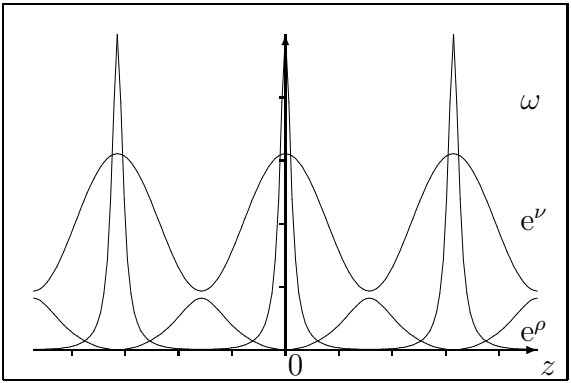


\unitlength=1mm


\caption{Gravitation fields distribution on axis $(r=0)$}\label{axis}
\end{figure}

Next class of solution may be constructed by using
the similar suggestion as in (\ref{PSclassII}).
\bear\label{ASclassII}
\psi=\const,&& \theta=\theta(\xi),\nn
\chi=\chi(\xi)&&\phi=\phi(\xi),\,\LAP\xi=0.
\ear
Set of ODEs in this case will be written as follows
\bear\label{ODU2ps}
\lal \ddot\theta+\frac{1}{2}\dot\chi^2\sh2\theta=0,\nnn
\ddot\chi+{2}\dot\chi\dot\theta\cth\theta=0,\\
\lal \ddot\phi+\frac{1}{2}\dot\theta^2+\frac{1}{4}\dot\chi^2\sh^2\theta=0.\nonumber
\ear
It's solution
\bear\label{FSas2}
\theta\eql\Arch\left\{k \sin\left|a(\xi-\xi_0)\right|\right\}\nn
\chi\eql\Arth\chi_0\pm\frac{a}{\sqrt{c^2-a^2}}\nn
\lal\times\arctg\left\{\displaystyle\pm\frac{c}
{\sqrt{c^2-a^2}}
\tg\left|a(\xi-xi_0)\right|\right\}\\
\phi\eql\phi_0+\phi_1\xi+\frac{1}{4}a^2\xi^2-
   á \displaystyle\int\limits^\xi\chi(\xi')d\xi',\nn
   k\eql\sqrt{1+c^2/a^2},\,a^2<c^2,\,c>0,\, a<0.\nonumber
\ear

The form of metrics coefficients will be the same as (\ref{MCas}), but
$F(\xi)$ and $G(\xi)$ will be determined in other way
\bear
F(\xi)\eql k \sin\left|a(\xi-\xi_0)\right|,\nn
G(\xi)\eql\frac{a}{c^2-a^2}\arctg\left\{\displaystyle\pm\frac{c}
{\sqrt{c^2-a^2}}
\tg\left|a(\xi-\xi_0)\right|\right\},\nn
\alpha^2\eql(1-\chi_0)^2 \nonumber
\ear

The solution existence conditions will look like discussed above.
\bear\label{condxiII}
\lal\frac{\arcsin k^{-1}+2\pi n}{|a|}\le |\xi(z,r)-\xi_0|<
\frac{\pi(2 n+\half)}{|a|},\nnn
\ch\chi(\xi)<\cth\theta(\xi).
\ear

Consider the method application in the case (\ref{ASclassI}-\ref{MCas}).
Let choice function-parameter in the form $\xi(z,r)=b (\e^{-r}\cos z+d)$.
Coefficients $b$ and $d$ are determined with respect to
of conditions (\ref{condxi}).

\begin{figure}[t]
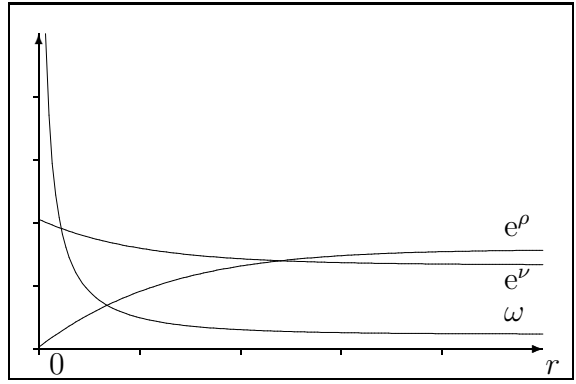

\unitlength=1mm

\caption{Gravitation fields distribution in plane $z=0$}\label{plane}
\end{figure}

Graphics of solutions, achieved by choosing $a=-1,c=1000,\xi_0=1,\chi_0=0.1$
are presented on figures \ref{axis} and \ref{plane}.

In the regions located along the axis $z$ gravitational fields distribution
possesses periodical character. In the plane, crossing axis on $z=0$ metric
coefficients rapidly decries to some constant value. Such a behaviour is
determined by the form of $\xi(z,r)$. Evidently, desire fields distribution
may be achieved by matching form of function-parameter.

\Acknow{
This work has been carried out under the aegis of the State
Scientific--Technical Program ``Astronomy. Fundamental Research of the
Cosmos,'' Section ``Cosmomicrophysics,'' and with partial financial support
from the Russian Foundation for Basic Research (Grant No. 98-02-18040) and
the Center of Cosmoparticle Physics ``Cosmion.''
Author thanks S.V.Chervon  and V.M.Zhuravlev for useful discussion.
}

\end{document}